\documentclass[doublecol]{epl2} 
\usepackage{graphicx}
\usepackage{epstopdf}
\title{Field-driven magnetisation steps in Ca$_3$Co$_2$O$_6$: A single-crystal neutron-diffraction study}
\shorttitle{Field-driven magnetisation steps in Ca$_3$Co$_2$O$_6$} 
\author{C.L.~Fleck,\inst{1} M.R.~Lees,\inst{1} S.~Agrestini,\inst{2,3} G.J.~McIntyre,\inst{4}  and O.A.~Petrenko\inst{1}}
\institute{  
\inst{1} Department of Physics, Warwick University, Coventry, CV4 7AL, UK\\
\inst{2} Laboratoire CRISMAT, CNRS-ENSICAEN, 14050 Caen, France\\
\inst{3} Max-Planck Institut CPfS, N\"othnitzer Str. 40, 01187 Dresden, Germany\\
\inst{4} Institut Laue-Langevin, 6 rue Jules Horowitz,  BP 156 - 38042 Grenoble Cedex 9, France}
\newcommand{\CCO}{$\rm Ca_3Co_2O_6$}
\newcommand\PRB[3]{Phys. Rev. B {\bf {#1}} ({#2}) {#3}}    
          
\newcommand\PRL[3]{Phys. Rev. Lett. {\bf {#1}}  ({#2}) {#3}}
\newcommand\JPSJ[3]{J. Phys. Soc. Japan {\bf {#1}} ({#2}) {#3}}


\abstract{
We present single-crystal neutron-diffraction data for the spin-chain compound \CCO. The intensity and line shapes of the two families of Bragg peaks characterising both the antiferromagnetic and the ferromagnetic components of the magnetic order present in this material have been measured as a function of temperature and applied magnetic fields of up to 5~T. We have studied the microscopic nature of the magnetic order at each step seen in the bulk magnetisation and investigated the evolution of the long and short-range components of the magnetic order in \CCO.}

\begin{document}
\maketitle
\section{Introduction} 
\CCO\ is one of a family of low-dimensional magnetic systems A'$_3$ABO$_6$, where A' is Ca or Sr, while A and B are transition-metal elements~\cite{Stitzer2001}.
It has a spin-chain structure ~\cite{Fjellvag1996} and undergoes a complex magnetic ordering below a N\'eel temperature of 25~K arising from the effects of geometrical frustration~\cite{Kageyama1997a,Kageyama1997b,Aasland1996,Kageyama1998,Hardy2003}.
In zero field, \CCO\ was recently shown to have a sinusoidally-modulated spin-density wave with a very long period propagating along the $c$ axis~\cite{Agrestini2008a,Agrestini2008b}.
A series of low temperature magnetisation steps~\cite{Kageyama1997a,Maignan2000,Hardy2004a}, reminiscent of the quantum tunnelling observed in molecular magnets~\cite{Maignan2004,Hardy2004b}, has ensured continued experimental and theoretical interest in this system in recent years ~\cite{Fresard2004,Wu2005,Kudasov2006,Kudasov2007,Kudasov2008,Yao2006a,Yao2006b,Soto2009,Qin2009,Chapon2009,Cheng2009}.
\begin{figure}[tb] 
\onefigure[width=0.95\columnwidth]{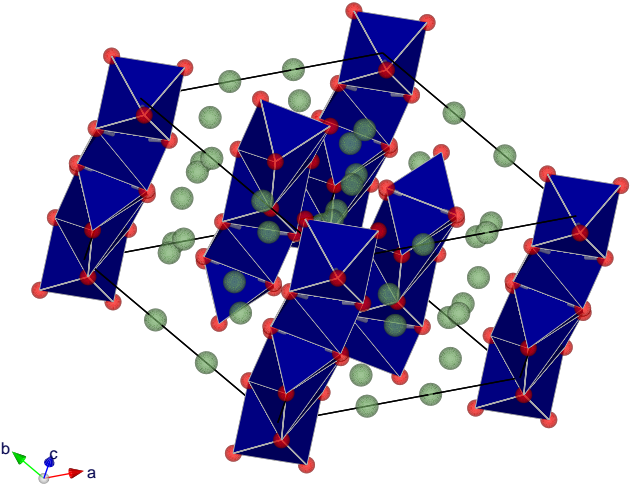}
\caption{The crystal structure of \CCO\ in the rhombohedral setting.
                The CoO$_6$ polyhedra form spin chains along the $c$ axis with the calcium atoms positioned between them.}
\label{Structure} 
\end{figure}

The crystal structure of \CCO~(space group $R\bar{3}c$)~\cite{Fjellvag1996} shown in figure~\ref{Structure} is highly anisotropic.
Along the $c$ axis there are chains of alternating CoO$_6$ trigonal prisms and CoO$_6$ octahedra with a Co-Co spacing of 2.59~\AA.
In the $ab$ plane the CoO$_6$ chains are arranged in a triangular lattice with a Co-Co separation of 5.24~\AA, double that along the chains~\cite{Fjellvag1996}.
Both the cobalt ions are trivalent with alternate high-spin $S=2$ and low-spin $S=0$ states corresponding to the CoO$_6$ trigonal prisms and the octahedra respectively~\cite{Aasland1996,Sampathkumaran2004,Takubo2005,Burnas2006}.
The intrachain coupling is  ferromagnetic (FM) while the interchain coupling is antiferromagnetic (AFM)~\cite{Kageyama1997a,Kageyama1997b,Aasland1996}.

At low-temperatures ($\leq 8$~K), a series of equally-spaced steps emerge in the magnetisation when a magnetic field is applied along the $c$ axis~\cite{Kageyama1997a}.
Magnetisation steps have been observed at $\sim 0$~T, 1.2~T, 2.4~T, 3.6~T, with possible steps at the higher fields of 4.8~T and 6~T~\cite{Maignan2000,Maignan2004}.
When the temperature is increased to an intermediate value ($\approx 10$~K), only the 3.6~T step remains and the hysteresis in the magnetisation as a function of magnetic field is greatly reduced.
At these temperatures, the magnetisation is saturated above 3.6~T with a value of $M_{sat}\approx5.1$~$\mu_B$/f.u, indicating complete ferromagnetic ordering.
Below 3.6~T the value of the magnetisation is $M_{sat}/3$.
This value is ascribed to two thirds of the chains having spin up and one third having spin down~\cite{Maignan2000}.
In a previous neutron diffraction study we have found a sharp change in the intensity of the magnetic Bragg peaks associated with the transition at 3.6~T. However, a small sample size and a low-flux instrument have prevented us from studying the transitions at any other fields~\cite{Petrenko2005}.
In this letter, we report the first observation of the meta-stable magnetisation plateaux in \CCO\ with a microscopic probe such as neutrons.

Another important aspect of behaviour of \CCO\ is a coexistence of short-range and long-range magnetic correlations~\cite{Agrestini2008b} and the associated decrease at low temperatures of the intensity of the main AFM Bragg peaks~\cite{Aasland1996,Kageyama1998,Petrenko2005}.
We also report on the balance between the short-range and long-range magnetic components in \CCO\ in an applied magnetic field.

\section{Experimental Details} 
An $8\times{2}\times{2}$~mm$^3$ single crystal grown using a flux method described previously~\cite{Agrestini2008b} was used for the experiment.
The high quality of the crystal was confirmed by x-ray diffraction, energy-dispersive x-ray, magnetisation, and specific-heat measurements.
The neutron diffraction measurements were made on the high-flux single-crystal diffractometer D10 at the Institut Laue-Langevin in Grenoble, France.

The sample was mounted in a 6~T vertical cryomagnet and aligned to within 1$^\circ$ of the magnetic field with the field directed along the $c$ axis limiting the scattering to the $(hk0)$ plane.
The incident wavelength used was 2.36~\AA\ from a pyrolytic graphite (PG) $(002)$ monochromator with the half-wavelength contamination suppressed by a PG filter.
Measurements were made in the temperature range 2-30~K.
The two-dimensional  $80 \times 80$~mm$^2$ area detector was used for all the measurements except for those made with the energy analyser, where a single $^3$He detector was used.

The measurements were made in one of two modes. Either by measuring the peak intensity (summing up the counts in a small area of the detector surrounding the peak) while ramping the temperature or field, or by 
integrating rocking curves (labelled as $\omega$ scans below) at various fixed values of field/temperature.
All the data have been normalised to the monitor time, typically 60~seconds per point in sweeping mode and 20~seconds per point for  $\omega$ scans.

The magnetic Bragg peaks observed in \CCO\ can be split into two groups, those appearing on top of nuclear peaks and labelled as FM peaks, and the AFM peaks, such as $(100)$, $(200)$ and $(120)$, whose intensity is zero above $T_N$.
In order to compare the intensity of the FM peaks with both the bulk magnetisation and the intensity of the AFM reflections, the nuclear component of the FM peaks was measured in zero field at $T>T_N$ and subtracted to give a purely magnetic intensity.

\section{Results and Discussion} 
\begin{figure}[bt] 
\onefigure[width=0.9\columnwidth]{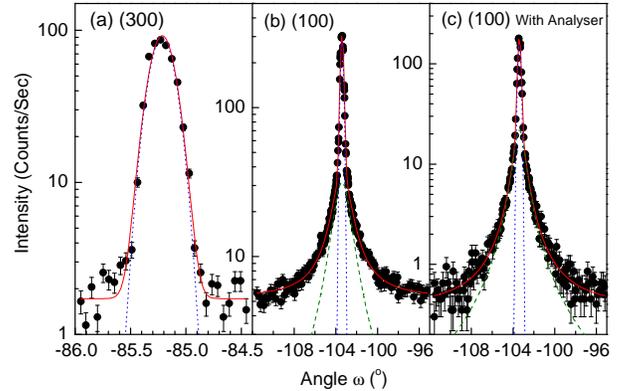} 
\caption{Examples of the lineshapes of the FM $(300)$ peak (a) and the AFM $(100)$ peak (b,c), measured at 2~K and in zero field.
                The profile shown in panel (c) was obtained with the use of a vertically focussing PG analyser.
                Solid lines are the fits consisting of two components, a main resolution-limited Gaussian component (dotted line) and a broad Lorentzian component (dashed line) as well as a flat background.
                The FM peak shown in panel~(a) can be fitted with a single Gaussian and a flat background.}
\label{Peak_Shape}
\end{figure}

We start the discussion of the results by comparing the characteristic shapes of the FM and the AFM reflections.
As illustrated in figure~\ref{Peak_Shape}a, the FM peaks, for example $(300)$, can be adequately fitted with a single Gaussian.
The width of these peaks (typical FWHM of $\simeq 0.35^\circ$) is considered to be limited by the instrumental resolution.

On the other hand, the AFM peaks, such as $(100)$, can only be fitted with a combination of a Gaussian and a Lorentzian component.
The Gaussian component (FWHM of 0.37$^\circ$) is again resolution limited, while the Lorentzian component is indicative of the presence of magnetic disorder (see figure~\ref{Peak_Shape}b).
This observation is in agreement with the previously reported coexistence of a short-range magnetic structure with the long-range order in zero field at 5~K~\cite{Agrestini2008b}.
The observed width of the Lorentzian component (typically 1.9$^\circ$) and the resulting estimate for the magnetic correlation length, $D=250$~\AA~\cite{width_calculation} are similar to the previously reported values ($D=180$~\AA\ was obtained in ref.~\cite{Agrestini2008b}).
 It is possible that this small difference is due to the truncation of the data in the earlier measurement, where typically the scans were much shorter than the 20$^\circ$ scans shown in figure~\ref{Peak_Shape}b and~\ref{Peak_Shape}c.

The AFM $(100)$ peak was also measured using the PG energy analyser with a typical peak profile shown in figure~\ref{Peak_Shape}c.
Apart from a narrowing of the Gaussian component due to the improved resolution down to 0.34$^\circ$ and a significant reduction in the background (from 5.3 to 0.4 counts per second) the shape of the peak remained largely unchanged.
The ratio of the areas of the Lorentzian to the Gaussian components of the peak  measured with and without the analyser are 0.7 and 0.9 respectively. 
This observation might suggest that the magnetic disorder is at least partially dynamic in nature.
However, to draw any firm conclusions regarding the presence of magnetic excitations in \CCO\ and their influence on the short-range magnetic correlations, much more systematic measurements with greatly improved energy resolution are required. 
\begin{figure}[tb] 
\onefigure[width=0.8\columnwidth]{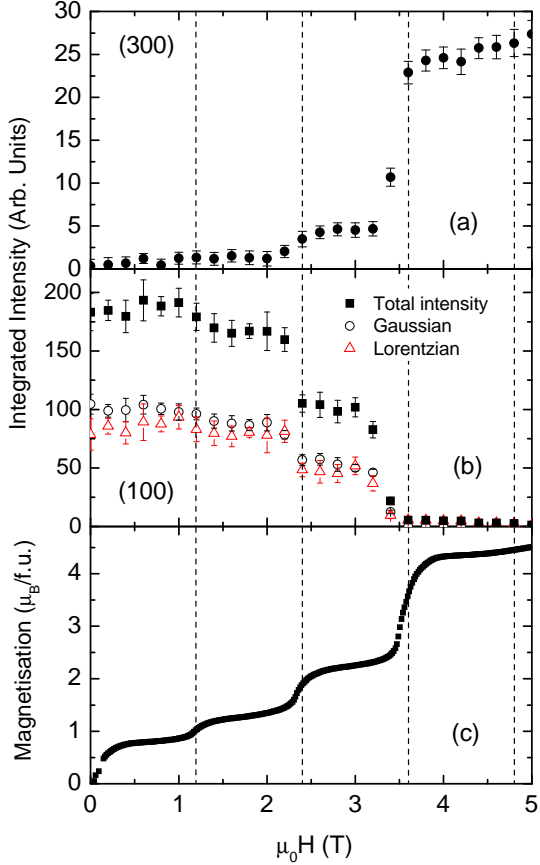}
\caption{Intensity of the FM $(300)$ reflection (a) and the AFM $(100)$ reflection (split into Gaussian and Lorentzian components) as a function of an increasing applied magnetic field.
               The nuclear component of the FM peak (equating to 24.9 in the units used in the figure) has been subtracted.
               The equivalent magnetisation measurements are shown for comparison (c).
               The dashed lines are drawn at fields of 1.2, 2.4, 3.6 and 4.8~T.
               All the measurements were made at a temperature of 2~K.}
\label{2K} 
\end{figure}

In agreement with our previous neutron diffraction study~\cite{Petrenko2005}, the application of a magnetic field did not result in the appearance of additional Bragg reflections.
All the peaks were found at integer positions which means that they can be indexed using the nuclear unit cell.
Figure~\ref{2K} shows the field dependence of the integrated intensity of the FM $(300)$ peak and the AFM $(100)$ peak, measured at 2~K with $\omega$ scans.
The magnetisation curve is also shown for comparison~\cite{MvsH}.
The equally-spaced steps in intensity for both the FM and the AFM reflections match well with those seen in magnetisation measurements, with clear steps observed at 2.4~T, 3.6~T and 4.8~T, and a possible feature at 1.2~T.
There is a small but distinctly nonzero intensity for the AFM reflection even above the transition at 3.6~T.
This field value is sufficient to polarise fully the system only at higher temperatures, while at 2~K a much higher field is required.

Figure~\ref{2K} also shows that the step-like behaviour is observed for both the Gaussian and the Lorenztian components of the AFM peak.
Therefore the application of a magnetic field affects both the long-range and the short-range antiferromagnetic correlations in \CCO.
The widths of both components of the AFM peak are found to be field independent until the transition at 3.6~T above which the width becomes resolution limited.

\begin{figure}[tb] 
\onefigure[width=0.99\columnwidth]{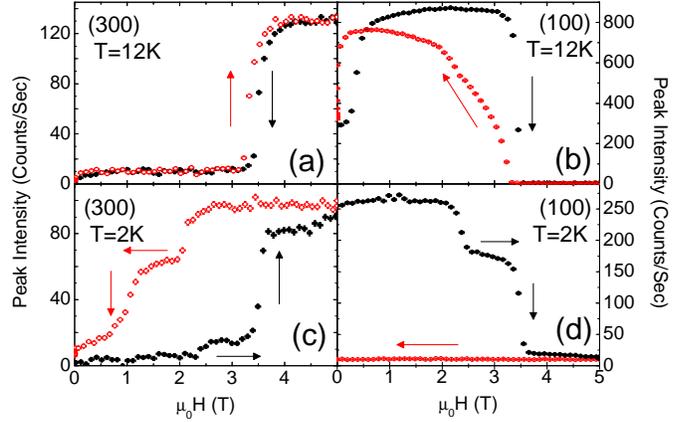}
\caption{The intensity of the FM peak $(300)$ peak (left) and the AFM peak $(100)$ (right) as a function of magnetic field at 12~K (a,b) and 2~K (c,d).
                The nuclear component of the FM peak (equating to 89 counts per second) has been subtracted.
                The solid symbols show the data taken while increasing the magnetic  field and the open symbols show the data taken while decreasing the field.
                The magnetic-field ramp rate was 0.1~T/min for all the measurements.}
\label{12and2KRamps} 
\end{figure}

The $\omega$ scans performed at various temperatures have shown that the positions of magnetic peaks in the $(hk0)$ scattering plane do not  change noticeably as a function of applied field.
It is therefore permissible to measure the field dependence of the intensity of the magnetic reflections without scanning through them.
The results of such measurements are shown in figure~\ref{12and2KRamps}.
The two different measurement methods give results that are in a good agreement with one other, as can be seen from the comparison between the field dependence of the intensity of an AFM peak $(100)$ and a FM peak $(300)$.
A similar step-like behaviour of intensity was also seen for the FM $(110)$ peak.

The field-ramping approach allowed much faster data collection, which is particularly important for \CCO, where the shape of the magnetisation curve is strongly dependent on the measurement time~\cite{Hardy2004a}.
The large hysteresis seen in the magnetisation measurements at low temperatures has also been observed with neutron diffraction, and is shown in figure~\ref{12and2KRamps}.
The most striking feature observed at base temperature is that the intensity of the AFM reflection is very weak (nearly zero) for decreasing magnetic fields.
Therefore at 2~K the ZFC magnetic state of \CCO\ is completely different from the magnetic state after the application and subsequent removal of an external magnetic field.
This observation is only possible with a microscopic magnetic probe such as neutrons while all the previous bulk-properties measurements have failed to note such a difference in the magnetic state after the application/removal of a magnetic field.
In order to restore the intensity of the AFM peaks the sample has to be warmed up to 30~K and then cooled in zero field.
\begin{figure}[tb] 
\onefigure[width=0.9\columnwidth]{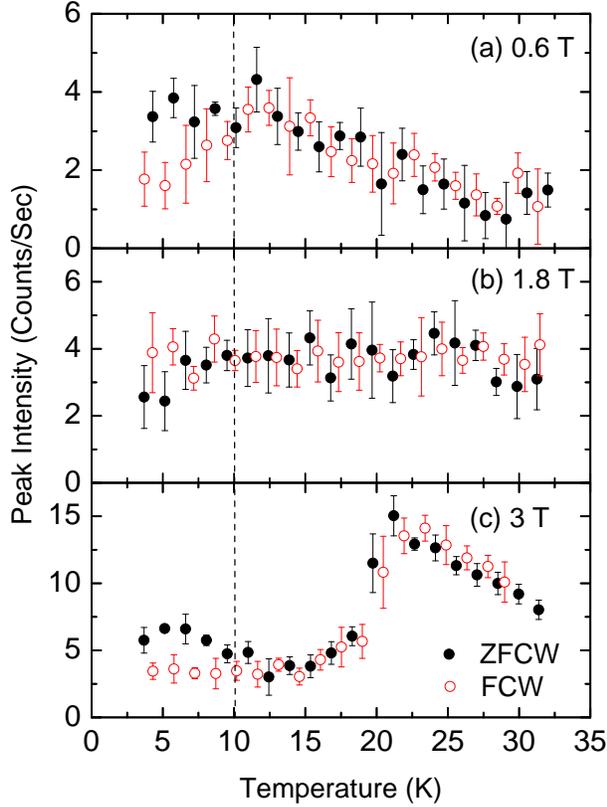}
\caption{Temperature dependence of the intensity of the FM $(300)$ peak measured in three different fields, 0.6~T, 1.8~T and 3~T.
                An irreversibility temperature is visible at around 10~K, where the intensities of the zero-field cooled warming (ZFCW) and field-cooled warming (FCW) diverge, marked by a dashed line on the figure.
                The temperature ramp rate was 0.3~K/min.}
\label{300Temp}
\end{figure}
  
One other peculiar feature observed in the measurements is the very small increase in the intensity of FM peaks close to 0~T.
This contrasts with the magnetisation results shown in figure~\ref{2K}c where a clear step in the magnetisation is observed close to 0~T  as the magnetic field is increased.
This is a particularly puzzling feature, as the absence of magnetic intensity in FM peaks such as the $(110)$ and the $(300)$ in zero field is considered to be a decisive argument in favour of a partially disordered antiferromagnetic (PDA) rather than a ferrimagnetic structure~\cite{Petrenko2005,Agrestini2008b}.

Unlike the FM peaks, the intensity of the $(100)$ AFM peak shows significant hysteresis at 12~K in a wide range of applied fields (see figure~\ref{12and2KRamps}b).
At this temperature there is also a rapid change in the intensity of the AFM reflection near zero field for the data taken in both ascending and descending magnetic fields,
while at 2~K the intensity of this peak does not change appreciably for ascending fields and is close to zero for descending fields.
The intensity curve for the AFM reflection taken while decreasing the field at 12~K suggests a transition at 2.4~T, which in magnetisation measurements~\cite{Hardy2004a} is also visible at temperatures as high as 10~K.
\begin{figure}[tb] 
\onefigure[width=0.95\columnwidth]{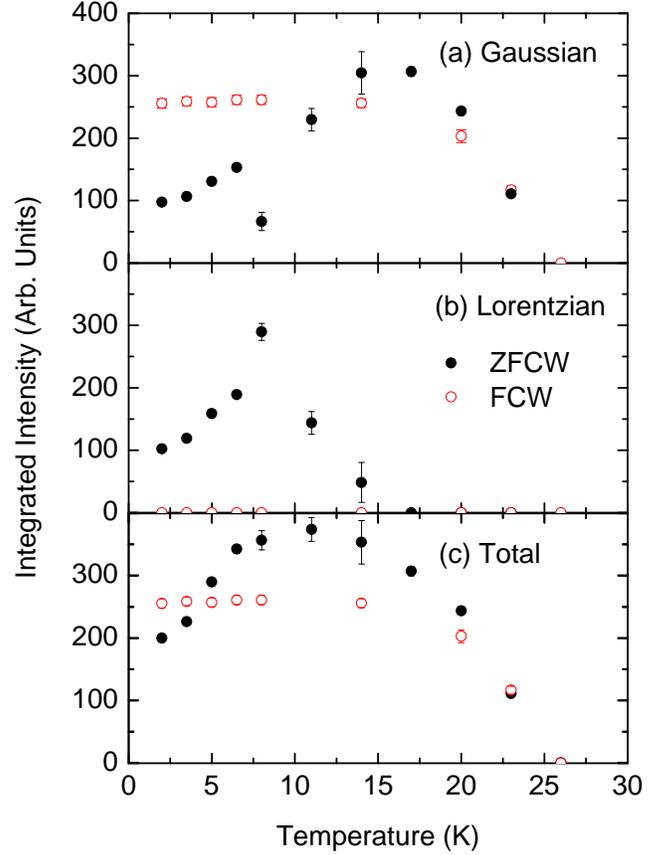}
\caption{The temperature dependence of the Gaussian and the Lorentzian components as well as of the total integrated intensity of the AFM $(100)$ peak in a field of 0.6~T.
               The solid symbols show the data taken while warming the sample in field after zero-field cooling (ZFCW) and the open symbols show the data taken in field while increasing the temperature after field-cooling (FCW).}
\label{100Temp0o6T} 
\end{figure}

The intensities of the FM peaks were also measured in magnetic fields of 0.6~T, 1.8~T and 3~T as a function of temperature.
These fields were chosen as they are the mid-points on the plateaux seen in the bulk magnetisation curve.
The results are shown in figure~\ref{300Temp}.
In addition to the apparent transition to the saturated state at 20~K in a magnetic field of  3~T (see figure~\ref{300Temp}c) another notable feature is the presence of an irreversibility temperature around 10~K.
Below this temperature there is a pronounced difference between the zero-field cooled and field-cooled data for 0.6~T and 3~T measurements, while in the intermediate field of 1.8~T the difference is barely visible.
This observation should be linked with the magnetisation relaxation measurements~\cite{Hardy2004a}, which revealed that at lower temperatures the magnetisation of \CCO\ has a pronounced time dependence and that the magnetisation could relax either upwards or downwards depending on the value of the applied field.

Recent investigations~\cite{Agrestini2008b} have suggested that the origin of the anomalous dip in the intensity of the AFM peaks at low temperatures is due to an increase in the fraction of material exhibiting short-range correlations at the expense of the fraction of long-range ordered material.
We have repeated the zero-field measurements reported in Ref.~\cite{Agrestini2008b} in a field of 0.6~T, the results of which are shown in figure~\ref{100Temp0o6T}.
For a ZFC sample both the Gaussian and Lorentzian components are present at base temperature.
On warming, the intensity of both components initially increases. There is a sharp dip in the Gaussian component at 8~K. However, above this temperature the Gaussian component continues to grow while the Lorentzian component diminishes rapidly.
The sharp decrease of the Gaussian component at 8~K indicates that some additional magnetic restructuring is taking place at this temperature, as the fraction of material exhibiting short-range order is at a maximum.
For a FC sample the peak shape is largely Gaussian and the short-range magnetic component is absent. At 2~K,  the overall intensity of the FC data collected in 0.6~T is higher that the ZFC intensity at the same temperature. The total intensity of the AFM (100) peak for the FC sample decreases monotonically as the temperature is increased and the ZFC and FC data sets crossover below the irreversibility temperature before coming together again at temperatures close to $T_N$.

Similar measurements made in fields of 1.8~T and 3~T (not shown) gave comparable results for the behaviour of the short-range and long-range components of the magnetic order in \CCO.

From inspection of figure~\ref{100Temp0o6T} it is obvious that, for a ZFC sample, the total intensity of the main AFM peak at base temperature is significantly lower than at 12~K.
Therefore, the presence of short-range correlations cannot fully account for the anomalous dip in the intensity of the AFM peaks at low temperatures as some intensity is still missing.

 \section{Conclusions}
The temperature and magnetic field dependence of both the FM and the AFM reflections in \CCO\ have been studied using single-crystal neutron-diffraction.
The use of a larger sample, and an instrument with a better resolution and a lower intrinsic background compared to previous measurements~\cite{Petrenko2005}, have enabled the first observation of changes in the intensity of magnetic peaks at each of the values of applied magnetic field where a step is seen in the bulk magnetisation.
The temperature and field dependences of the FM peaks mirror that previously reported in the bulk magnetisation data. At 2~K, the application of magnetic fields of up to 5~T  does not lead to the complete disappearance of the intensity on the AFM peaks and the intensity on these peaks is not restored by the subsequent removal of the field. The magnetic field dependence of the short-range correlations in \CCO, which in this study are reflected in the line shape and intensity of the AFM peaks,  has also been measured for the first time. Both components decrease in a stepwise fashion as the magnetic field is increased. The increase in volume of the material with short-range order cannot completely account for the loss in intensity from the AFM peaks seen as the temperature is reduced well below $T_N$.
\acknowledgments
The authors would like to thank the EPSRC for funding.
The VESTA software package~\cite{VESTA} was used in the drawing of figure~\ref{Structure}.


\begin{thebibliography}{99}
\bibitem{width_calculation} The magnetic correlation length is estimated using D(\AA)=2$\pi$/FWHM(\AA$^{-1}$).
\end{thebibliography}
\end{document}